\documentclass[12pt]{article}
 
\usepackage{amsmath,amssymb,dsfont,mathrsfs,graphicx, float}
\usepackage[hmargin=1.75cm,vmargin=1.65cm]{geometry}

\begin{document}

\title{Closed \,\, Form \,\, Solutions \,\, To \,\, Bosonic \,\, Perturbations \,\, In \,\, General \,\, Relativity}
\author{Arthur Suvorov$\footnote{Arthur.Suvorov@monash.edu}$, Anthony W.C. Lun$\footnote{Anthony.Lun@monash.edu}$}

\date{January, 2014}

\date{Monash Center for Astrophysics, School of Mathematical Sciences, Monash University, Wellington Road, Melbourne 3800, Australia}

\maketitle

\Large
\begin{center}
{$\mathbf{Abstract:}$}
\end{center}
\bigskip
\normalsize
We present some results regarding metric perturbations in general relativity and other metric theories of gravity. In particular, using the Newman Penrose variables, we write down and discuss the equations which govern tensor field perturbations of ranks $0, \pm 1$ and $\pm 2$ (scalar,vector,tensor bosonic perturbations) over certain space-times that admit specific background metrics expressible in isotropic coordinates. Armed with these equations, we are able to develop the Hadamard series which can be associated with the fundamental solution of the equations, wherein we introduce an inhomogeneous singularity at the physical space-time point of the perturbing particle. The Hadamard series is then used to generate closed form solutions by making choices for an appropriate ansatz solution. In particular, we solve for the spin-weighted electrostatic potential for the Reissner-Nordstrom black hole and for the fully dynamical potential for the Friedmann-Robertson-Walker cosmological solution.

\section{Introduction}
We investigate closed form solutions to bosonic perturbation equations in the theory of general relativity. Bosonic particles represent particles present in the standard model that have integer spin $\sigma$, $\sigma \in \mathbb{Z}$. They therefore encompass scalar particles (such as the Higgs boson), vector particles (such as the photon and classical electron) and tensor particles, which are arguably the most interesting, and take the form of the graviton. The resulting perturbation equations are very naturally useful, since they describe, to leading order, the dynamics of these particles as they encounter various astrophysical objects. The exact nature of the astrophysical objects which we are interested in is encoded within the metric tensor of the space-time which will be a solution of the background Einstein-field equations with some potentially non-zero matter Lagrangian minimally coupled via the Levi-Cevita connection: $G_{\mu \nu} =  8 \pi T_{\mu \nu}$\footnote{Note throughout this paper we assume natural units $c = G = \hbar = 1$.}.
\\
This type of work was instigated by E.T. Copson in 1928 [4], but has since apparently lost its value within the literature. Copson managed to develop a closed form solution to the electrostatic problem around a Schwarzschild black hole, and has yielded fruitful results in the context of self force calculations [5]. We should like to reignite Copson's and others' efforts who have taken on his approach by demonstrating the power of his methods.
\\
The first goal of this paper is to extend the apparent usefulness of the Hadamard method to incorporate more general perturbation equations than those currently in the literature, i.e, to all massless bosonic perturbations rather than just electromagnetic and scalar. In particular, we wish to find fully closed form electrostatic potentials associated with black hole background metrics. This possibility has been explored to a great extent by Linet and others [26,27], who shows that perturbations around black holes in general relativity admit such electrostatic solutions for a wide range of background metrics. We extend Linet's results briefly for stellar interior solutions also in the case of a spin-1 perturbation.
\\
The second main goal here is to show that for any space-time in which the Weyl tensor $C^{\alpha}_{\beta \gamma \delta}$ vanishes, thereby permitting a coordinate system where the metric is locally flat, i.e $g_{\mu \nu} =  e^{2 \psi} \eta_{\mu \nu}$ [32], closed form solutions exist to the fully dynamical perturbations under certain algebraic conditions.
\\
In light of no-hair and uniqueness theorems (such as Israel's theorem), the black hole we will be interested in primarily is the Reissner-Nordstrom black hole, since it reduces to the Schwarzschild solution in the limit $Q \mapsto 0$ and admits desirable algebraic properties as a space-time for the purposes of perturbation analysis [28].
\\
In summary, the goal of this paper is to extend the Hadamard-Copson-Linet approach to two major classes of cases, wherein we find and analyse closed form solutions for:
\begin{itemize}
\item Electrostatics for bosonic perturbations of Reissner-Nordstrom black holes;
\item Fully dynamical potential functions for bosonic perturbations in a space-time that admits vanishing Weyl tensor $C^{\alpha}_{\beta \gamma \delta}$ (Cosmology).
\end{itemize}
For a review on these types of perturbations one can consult $[14]$ or $[13]$ for a more cosmological treatise.
\\
This paper is organised as follows: In section II we describe the elements of general relativity essential to our construction and the soultions to the field equations we are interested in. In section III we express the relevant perturbation equations in terms of a 'master' differential operator and develop the Hadamard theory that accompanies that equation. Sections IV and V are devoted to using the symmetries encoded within the Hadamard expansions to reduce the partial differential equations (PDE) from section III to ordinary differential equations (ODE). Finally, in the last sections we solve these ODE and make some comments regarding the nature of these solutions. In addition to these results, we also provide a closed form solution for an electrostatic perturbation of spins 0 and 1 of the Reissner-Nordstrom-Brans-Dicke black hole in Appendix A and the closed form solution of a spin 1 perturbation inside a stellar interior in Appendix B.
\subsection{Assumptions And Scope}
The Hadamard-Copson-Linet approach to solving electrostatic problems in curved space-times is not restricted to any particular metric theory of gravity [26]. It is for this reason that we consider a very general class of metrics, namely those which belong to the Einstein-Euler-Maxwell  theory. However, this approach is not mathematically limited to a specific metric theory, so we stress its universality. Nevertheless, for the purposes of understanding $\mathit{all}$ bosonic perturbations, we begin with the theory of general relativity. This approach is taken so that the linearised Einstein equations in vacuum $G^{B}_{\alpha \beta} = 0$ admit a manageable form and have many nice properties proven in the literature that we will make use of (see e.g [28,29]). The theory here can be completely generalised to $\mathit{any}$ metric theory of gravity for the spin $0$ and spin $1$ cases, and we will include at the end the solutions for a scalar and electrostatic perturbation to a Brans-Dicke black hole to demonstrate. 
\\
We will use a rather general metric tensor for the most part, where we assume a line element with isotropic spatial component:
\begin{equation}
ds^2 = -e^{2 \alpha(r,t)} dt^2 + e^{2 \beta(r,t)}(\delta_{ij} dx^{i} dx^{j})
\end{equation}
For existence conditions on such an isotropic coordinate system one may consult [22]. We therefore restrict ourselves to the cases where a coordinate transformation permits the background metric for the space-time to be cast in the form (1).
\\
Note, when we consider black-hole space-times within the framework of general relativity we may assume the metric functions simplify: $\alpha(r,t) = \alpha(r)$ and $\beta(r,t) = \beta(r)$. This is due to the fact that a Birkhoff theorem applies for the Einstein equations and also for the Einstein-Maxwell equations [21]. Any metric theory wherein a Birkhoff theorem applies can also have this reduction applied. Therefore, we will separate the cases later on to simplify the algebraic manipulations. 
\\
Now, having written down our general metric (1), we wish to examine the nature of the equations which will govern a spin-$\sigma$ perturbation for $\sigma \in \mathbb{Z}$. This will be accomplished by the well established methods of Newman and Penrose, wherein we express the metric tensor in terms of tetrad components [23]. 

\subsection{Motivation}
Understanding metric perturbations is a very important part of linear analysis in general relativity and more general metric theories of gravity. These metric perturbations answer questions of: stability, dynamics, evolution and signal processing. In essence, the metric perturbations allow one to probe the background space-time from which we receive signals from. Having a deep understanding of the nature of the space-time geometry allows us to formally conclude the relative signal frequencies and amplitudes that astrophysical objects emit constantly [36]. This work has been conducted since the inception of general relativity, and in particular linear analysis has played a huge role [14,25,11]. 
\\
We stress here what our paper has to offer. For not only physically interesting but mathematically interesting reasons, we derive new solutions to the perturbation equations which govern the metric perturbations of a background space-time by a bosonic particle. In the literature, it is customary to decompose the partial differential equations which describe the phenomena we are interested in as a separation of variables and to then solve the equations as a set of ordinary differential equations resulting in a multipole solution. While this does in fact yield a solution, authors seldom discuss the limitations of this approach. Lacking existence and uniqueness results, which is the case for a lot of these perturbation equations, the separation of variables approach may in fact constrict the solution space, and provide merely a glimpse of the actuality. This is the mathematically inclined perspective, wherein we find closed form solutions which do not obviously relate to their series counterparts. Furthermore, even if the equations do admit a unique solution, one can then match the separated solution to the closed form one and obtain new summation formulae [5]. We stress here that the idea of finding closed form solutions is not entirely new, and this research has been conducted by a few already [26,27,5,24]. Our approach generalises theirs to a certain extent.
\\
For the physically motivating approach, one can imagine that meaningful analysis can be conducted much easier with a closed form solution. Indeed, interesting effects such the existence of Meissner type effects [3] can be investigated. Low energy limit analysis also becomes much simpler, so one can investigate things such as Newtonian limits of the perturbation variables [1]. Furthermore, for the purposes of self-force calculations, a closed form solution to a general bosonic perturbation makes the next steps much simpler. We leave these extra details to future work.

\section{General Relativity: Essentials}
We discuss here a few components of general relativity that will be particularly important to the following work. We have that the field equations for the Einstein-(Maxwell-Euler) system are given by:
\begin{equation}
G_{\alpha \beta} = 8 \pi T^{Matter}_{\alpha \beta}
\end{equation}
Where $G_{\mu \nu}$ is the Einstein tensor and $T_{\mu \nu}$ is the stress energy tensor generated by an electromagnetic or fluid Lagrangian that takes the form of the quadratic Faraday tensor or of the Euler Lagrangian.
\\
The field equations (2) will allow us to write down the Regge-Wheeler operator corresponding to a linearised perturbation about the background space-time which satisfies these exact field equations. One may consult [30] for a linearised treatment of general relativity.

\subsection{Black-Holes $\&$ Relativistic Stars}
Within this section we discuss the black hole and relativistic star solutions which will be of interest to us in our calculations. In particular, the black hole we will analyse will be the Reissner-Nordstrom solution. The line element for this metric is given as, in Boyer-Lindquist coordinates [18]:
\begin{equation}
ds_{RN}^2 = -f(r) dt^2 + \frac {1} {f(r)} dr^2 + r^2 d \Omega^2 
\end{equation}
with $f(r) = 1 - \frac {2 M} {r} -  \frac {Q} {r^2}$. Here $M$ is the usual mass parameter and $Q$ is the charge parameter of the solution, which can be defined as the momenta of the metric [18].
\\
As discussed already in the previous sections we will require an isotropic coordinate transformation, this can be accomplished with $r \mapsto \tilde{r} \big( (1 + \frac {M} {2 \tilde{r}})^{2} - (\frac {Q} {2 \tilde{r}})^{2} \big)$ and yields:
\begin{equation}
ds_{RN}^2 = - \frac {(\frac {r-B} {r+B} )^{2}} {\eta(r)^2} + \eta(r)^2(1+ \frac {B} {r})^{4} (dr^2 + r^2 d \Omega)^2 \,\,\,\, ; \,\,\,\,\, \eta(r) = p_{+}^2 - p_{-}^2 (\frac {r-B} {r+B} )^{2}
\end{equation}
Where we have defined $p_{+}^2 = \frac {M + \sqrt{M^2 - Q^2}} {2 \sqrt{M^2 - Q^2}}, p_{-}^2 = \frac {M - \sqrt{M^2 - Q^2}} {2 \sqrt{M^2 - Q^2}}$ and $B = \frac {1} {2} \sqrt{M^2 - Q^2}$.
\\
In matching expression (4) with (1) we can easily choose appropriate $\alpha$ and $\beta$ metric coefficients. This black hole will be one of the focuses of this paper, and we will derive closed form solutions to a bosonic perturbation with respect to this background metric. We note here, again, that we can thus far only solve the bosonic electrostatic problem associated with the metric (4). This is a good first step however, and provides interesting analysis nevertheless, e.g in the context of a 3+1 ADM split. Using separation of variables, a fully dynamical solution can be found. Or, failing that, one can employ a Fourier transform for the potential variable:
\begin{equation}
V(t, \mathbf{x}) = \int^{\infty}_{-\infty} \tilde{V}(\omega, \mathbf{x}) e^{- i \omega t} d \omega
\end{equation}
in an attempt to obtain a fully dynamical solution in Fourier space.
\\
The next focus would be in solving electrostatic problems associated with the interior solutions of a general star. Indeed, these are typically found to be the solutions of the Tollman-Oppenheimer-Volkoff equation [31]. We will here express a few particular versions of this, in particular an analogue of the Vaidya-Tikekar solution [1,6,7]. The line element for this solution can be written as, in isotropic coordinates:
\begin{equation}
ds_{VT}^2 = \frac {4} {(1 + \alpha r^2)^2} [(-\beta^2 + \alpha r^2)^2 dt^2 + dr^2 + r^2 d \Omega^2]
\end{equation}
Here the parameters $\alpha$ and $\beta$ relate to the relative strengths of the pressure and density respectively.
\\
As it has been shown, the above solution can be smoothly matched to the Reissner-Nordstrom exterior (4). For a nice review on stellar interior solutions one could consult [31]. As of yet we are only able to solve the spin-1 case for this metric, but it seems hopeful that we can extend this result to incorporate an arbitrary bosonic perturbation. Results for the closed form electrostatic solution for this metric can be found in Appendix B.

\subsection{Cosmology}
We discuss here, very briefly, the cosmological solutions we attempt to analyse. Indeed, we will attempt to solve the fully dynamical metric perturbations for space-times that admit vanishing Weyl tensor $C^{\alpha}_{\beta \gamma \rho} = 0$. As has been discussed already, such solutions admit coordinate systems wherein the metric can be cast in a Jordan frame to make it locally flat, i.e to be conformally flat. Such solutions of the Einstein-Euler field equations (2) are very interesting cosmologically [34]. We will not discuss here the physical significance of certain cases, but merely exploit the Hadamard method to generate solutions to these perturbation equations. One can readily imagine why it would be interesting to understand dynamical behaviour of introduced gravitons inserted into a cosmological space-time solution. In particular, we will be interested in Friedmann-Robertson-Walker cosmology, which falls into this class [12]. Using the Friedmann equations expressed in a conformally flat coordinate system, one can solve the fully dynamical perturbation equations in closed form, as we shall see.

\section{Spin-Weighted Wave Equations}
We begin by writing down the spin-$\sigma$ perturbation equation for spin parameter $\sigma = 0, \pm 1,\pm 2$ in the local isotropic coordinate basis defined in section I. Let $u$ be an arbitrary vector field living on the space-time manifold $u \in \mathfrak{X}(M)$, and $\mathcal{L}_{u}$ be the Lie derivative taken along $u$. Using linear perturbation theory and the fact that the perturbed (linearised) Bianchi identities are independent from vector flows: 
\begin{equation}
\mathcal{L}_{u} R^{B}_{i [jk \ell]} = 0
\end{equation}
it is possible to derive a single 'master' wave equation describing the propagation of perturbing bosonic fields, such as the Penrose wave equation and Teukolsky master equation [10]. The fact that these equations are invariant under this Lie derivative makes them particularly valuable, as we can be certain that they describe physically meaningful and unambiguous phenomena because they are then gauge invariant [9,34]. This means that we may then $\mathit{choose}$ a gauge which simplifies the calculations. The gauges which we employ will depend on the tensorial nature of the perturbation and will be either the Lorentz gauge or Regge-Wheeler gauge. Following in the path of [23,25,28], we express the generalised Regge-Wheeler operator $\mathcal{R}_{\sigma}$ in terms of the basis elements in the Newman-Penrose formalism:
\begin{equation}
\mathcal{R}_{\sigma} = (D+ 2 \epsilon - (1 + \sigma)\rho)(\Delta + (1 + \sigma)\mu) - (\delta + 2 \beta) \bar{\delta} - (1 - \sigma^2) \Psi_{2}
\end{equation}
Where the elements are given by their standard definitions (see e.g [23]). Note that $\Psi_{2}$ here is one of the Weyl scalars. For a review and thorough derivation one can consult [29,8,15].
\\
Note that, when we take the spin parameter $\sigma$ equal to $0$, definition (8) reduces to the covariant Klein-Gordon operator: $\mathcal{R}_{0} =  g^{\mu \nu} \nabla_{\mu} \nabla_{\nu}$ which generates the equation for a scalar perturbation. Similarly,  the Maxwell operator: $\mathcal{R}_{1} = \partial_{\mu} (\sqrt{-g} g^{\mu \nu} g^{00} \partial_{\nu} A_{0})$ is recovered for $\sigma = 1$ after one has exploited the Lorentz gauge $\nabla^{\mu} A_{\mu} = 0$. Furthermore, the classical Regge-Wheeler operator (linearised Einstein-equation): $\mathcal{R}_{2} = G^{B}_{\alpha \beta} V^{\alpha \beta}$ is too when we set $\sigma = 2$ and exploit the Regge-Wheeler gauge conditions [10,19]. In essence, equation (8) may be decomposed into three separate equations, each of which represents an integer spin perturbation. Note that the input function that we act on with the operator $\mathcal{R}_{\sigma}$ is always ultimately a scalar function representing the potential associated with the wave packet of the perturbation. As such, the wave packets generated by the potential are scattered against the potential barrier and cause an interference with the dynamics.
\newline
Employing the local coordinate basis $\{x^{\mu}\}_{iso}$ described by the line element (1) and lifting these local element expressions to the tangent bundle, we are able to express the operator $\mathcal{R}_{\sigma}$ as:
\begin{equation}
\mathcal{R}_{\sigma} = -e^{-2 \alpha} (\partial_{tt} + [-(1 + 2 \sigma)\partial_{t} \alpha + 3 \partial_{t} \beta] \partial_{t}) + e^{-2 \beta} (\nabla^{2} + [(1-2 \sigma) \partial_{r} \alpha +  \partial_{r} \beta] \partial_{r})
\end{equation}
Where $\nabla^{2}$ is the standard Laplace operator, which, in Euclidean coordinates, assumes the expression $\nabla^{2} = \delta^{ij} \partial_{i} \partial_{j}$.
\\
Assuming a single charge to be the source of the perturbation, we are able to write down the explicit perturbation equation as:
\begin{equation}
\mathcal{R}_{\sigma} V(t,r,\theta,\phi) = J^{0}
\end{equation}
Where $J^{0}$ is the current density associated with the perturbation and is conformally related to the Dirac measure $\delta(x^{i} - x^{i}_{0})$. The charge $J^{0}$ exists to ensure that the solution to equation (10) takes the form of a $\mathit{fundamental}$ solution, since the right hand side is discontinuous at the point wherein we introduce the perturbing particle, i.e $x^{\mu}_{0}$. This is to comply with physical expectations, and a more rigorous explanation can be found in [5] where the intruding point-like tensorial charge assumes a density profile of the form:
\begin{equation}
J^{0} = 4 \pi \rho(t, \mathbf{x}) = \int^{\infty}_{-\infty} (1/\sqrt{-g}) \delta^{4} (x^{\alpha} - b^{\alpha}(\tau)) d \tau
\end{equation}
\\
Where $b^{\alpha}$ is the spacetime trajectory of the particle.
\\
Let us make some remarks here regarding the scalar nature of equation (10). One familiar with metric perturbation theory will be aware that a tensorial perturbation of rank $n$ ought to result in a tensor equation of rank $n$. However, using gauge choices such as the Lorentz gauge condition $\nabla_{\mu} A^{\mu} = 0$ for spin-1 dynamics, one is able to eliminate all tensorial variables except for one. The remaining components of the tensorial perturbation can then be deduced by backtracking the gauge condition and solving for the other components of the electromagnetic potential $A_{\mu}$ for vector perturbations or the components of the metric tensor $h_{ij}$. This metric tensor takes its form from $g_{\alpha \beta} = \eta_{\alpha \beta} + h_{\alpha \beta}$ wherein one performs a linear perturbation. For a concise explanation of this gauge invariance cleverness one can consult [2,10]. It is because of this that we have forsaken no generality in our approach.
\\
The goal of this paper is to, therefore, solve equation (10) without assuming a separation of variables for a few different classes of metric coefficients choices as outlined in the introduction. The main tool by which we will analyse and decompose equation (10) will be via the Hadamard method, which we now go on to discuss
\subsection{Hadamard Method}
We begin this section with a theorem of Hadamard, which puts our approach on a nice footing:
\\
$\mathbf{Theorem \,\,\, (Hadamard)}$:  Consider a second-order linear partial differential equation of the form:
\begin{equation}
\mathcal{D}u = \eta^{ij} \frac {\partial^2 u} {\partial x^{i} \partial x^{j}} + h^{j} (r,t) \frac {\partial u} {\partial x^{j}} = 0
\end{equation}
where we assume the $h^{j}$ are differentiable functions of the radius $r = \delta^{ij} x_{i} x_{j}$ and the temporal coordinate $t$. The fundamental solution to equation (12) is continuous and differentiable everywhere except possibly at a singular point $(t,r,\theta,\phi) = x^{\mu}_{0} = (t_{0},b,\theta_{0},\phi_{0})$ and can be written as:
\begin{equation}
u = \frac {U(t,r,\theta,\phi)} {\Gamma^{m}}
\end{equation}
where $\Gamma$ is given as the square of the Minkowski distance from the perturbation point: $\Gamma = \eta_{\mu \nu} X^{\mu} X^{\nu}$ where we have $X^{\mu} = x^{\mu} - x_{0}^{\mu}$ and $m \in \mathbb{Q}$. Furthermore, $U$ is expandable as a convergent power series in $\Gamma$ such that:
\begin{equation}
U(t,r,\theta,\phi) = \sum_{j} U_{j} \Gamma^{j}
\end{equation}
\\
Where each $U_{j}$ is an analytic function of the coordinates. Note the above expansion (14) applies when the problem is $\mathit{odd}$ dimensional, otherwise there $\mathit{may}$ be an additional series involving powers of $\log{\Gamma}$. 
\\
A proof of this theorem can be found due to Hadamard in his original works [16]. One will now note that our equation (10) is of this form away from the perturbation point $x^{\mu}_{0}$ provided we assume $\phi$-symmetry.
\\
The $\Gamma$ here form an orthogonal basis for a Hilbert space, which should make it clear why the approach yields a manageable set of equations. Indeed, one assumes orthogonality between the $\Gamma$ and $U$ functions and as such will obtain a set of equations of the form:
\begin{equation}
\Gamma^{i} (\mathcal{D}^{0} U_{\sigma,n}) + \Gamma^{i+1} (\mathcal{D}^{1} U_{\sigma,n}) + \Gamma^{i+2} \cdots = 0
\end{equation}
For some $i$ and differential operators $\mathcal{D}^{j}$. One can then solve the system of equations:
\begin{equation}
\mathcal{D}^{j} U_{\sigma, n} = 0 \,\,\,\,\, \forall j
\end{equation}
to obtain recurrence relations for the $U_{\sigma, n}$. This is the essence of the Hadamard approach.
\\
The above theorem assures us that there will exist a Hadamard expansion for our perturbation equations. The hope is then that, using symmetry insight from the Einstein field equations, one can deduce a clever ansatz for the equation (10). Note further that, for the case of electrostatic perturbations, the expression for the Minkowski distance $\Gamma$ reduces to the 3-dimensional Euclidean version: $\Gamma^{static}_{M^{4}} = \Gamma_{\mathbf{E}^3}$ due to the absence of the time coordinate.
Omitting the tedious algebraic calculations, let us now assume that the potential function takes the form:
\begin{equation}
V = \sum_{n \in \mathbb{N} \cup \{0\}} U_{\sigma,n} \Gamma^{n - \frac {1} {2}}
\end{equation}
Acting on this expression $V$ and assuming analyticity conditions to pull in the differential operators under the summation sign we are able to find that the Hadamard coefficients $U_{\sigma,n}$ obey the relation (assuming $\Gamma$ orthogonality):
\begin{equation}
\sum_{n=0}^{\infty} (2n-1) \Gamma^{n - \frac {3} {2}} \Big[ \frac {U_{\sigma,0}} {r} (\Gamma + r^2 - b^2) \frac {d \big( \frac {U_{\sigma,n}} {U_{\sigma,0}} \big)} {d r} + 2n U_{\sigma,n}  + \frac {1} {2 n -1} \mathcal{R}_{\sigma} U_{\sigma,n-1} \Big] = 0
\end{equation}
Where $U_{\sigma,0} = e^{- \int^{r}_{b} \frac {-h_{\sigma}(r)} {2} dr}$ is found to be the first term in the series.
Solving equation (18) recurrently we have:
\begin{equation}
\frac {U_{\sigma,0}} {r} (2n-1) (r^2 - b^2)^{1 -n} \frac {d} {d r} \big[ (r^2-b^2)^{n} \frac {U_{\sigma,n}} {U_{\sigma,0}} \big] = -\mathcal{R}_{\sigma} U_{\sigma,n-1} - \frac {U_{0}} {r} (2n -3) \frac {d} {d r} \big( \frac {U_{\sigma,n-1}} {U_{\sigma,0}} \big)
\end{equation}
We can therefore conclude that:
\begin{equation}
\frac {U_{\sigma,n}} {U_{\sigma,0}} = \frac {(r^2-b^2)^{-n}} {2n -1} \int^{r}_{b} \frac {r(r^2 -b^2)^{n-1}} {U_{\sigma,0}} \Big[ -\mathcal{R}_{\sigma} U_{\sigma,n-1} - \frac {U_{\sigma,0} (2n -3)} {r} \frac {d} {d r} \big( \frac {U_{\sigma,n-1}} {U_{\sigma,0}} \big) \Big] dr
\end{equation}
Where we have used the boundary condition $U_{\sigma,n}(b) = 0$ for $n > 0$ and defined $U_{\sigma,n} = 0$ for $n < 0$. Expression (20) allows us to compute any number of $U_{\sigma,n}$ to the desired order in the expansion, thereby solving equation (10) [17]. However, this still leaves us an infinite series solution which is not desirable. As such, we will extract important information from expression (20) and deduce which metrics admit closed form solutions in this Hadamard sense. Similar expressions were obtained by Wiseman in his paper for scalar perturbations [5] and our result reduces to his as $\sigma \mapsto 0$.
\\
\section{Spin-Weighted Electrostatics}
We first discuss the ansatz functions we will choose taking insight from the Hadamard series foliation.
\\
Indeed let us make use of the substitutions:
\begin{equation}
V(r,\theta) = g(r) F(\gamma)
\end{equation}
\begin{equation}
\gamma(r,\theta) =  \frac {\Gamma(r,\theta)} {f(r) f(b)}
\end{equation}
\begin{equation}
g(r) = \frac {U_{\sigma,0}} {\sqrt{f(r)}}
\end{equation}
\begin{equation}
f(r) = \frac {r^2} {B} e^{\alpha(r) + \beta(r)}
\end{equation}
These forms will, as we shall see, allow one to reduce equation (10) to an ordinary differential equation, which can then in principle be solved. The form of these expressions may seem mysterious, but they are really not. Indeed, in physics one often makes use of symmetry principles to reduce the degrees of freedom within equations. Here we employ that exact ideology, and the above functions are chosen so that they encode the symmetry properties of the space-time metric (1). The Hadamard series foliates in a way describing the Green's function associated to a differential operator [17], from which the potential can then be built. The $U_{\sigma,0}$ is the leading order term in the series, and as such plays the main role. Nevertheless, this function $f$ here, while apparently somewhat cryptic in nature, actually is a result of the foliation of the right hand side of equation (19) once the Einstein field equations have been used. The details are somewhat lengthy, but one can imagine that the Einstein field equations for the metric (1) necessarily demand relationships between the metric coefficients, which then makes demands on the functional form of the Regge-Wheeler operator. These demands translate to simplifications of the RHS of (20) and in turn give simplified expressions under the integral present. The relevant information can then be extracted, and indeed takes the form of (24).
\\
Note that, the functional form for $f$ above only applies for Black-Hole space-times. For stellar interiors and cosmological space-times the form for $f$ differs, and is given as, explicitly:
\begin{equation}
f_{Int}(r) = \frac { \mu r^2 - \kappa^2} {\iota} \,\,\,\,\, ; \,\,\,\,\, f_{Cos}(t) = \frac {1} {\kappa^2 (t-t_{0})^2}
\end{equation}
For some appropriate choice of constants $\kappa, \iota$ and $\mu$. This is because the $B$ parameter relates to the horizon structure of the black hole, which is ill defined for different space-times.
\section{Differential Equation Reductions and Solutions}
In this section we reduce the differential equation (10) by employing the important characteristics of the Hadamard method and then proceed to solve them.

\subsection{Black Holes}
In this section here we will solve the electrostatic problems for the metrics outlined in section II.I, that is for the Reissner-Nordstrom exterior solution.
Here we will assume a static solution to equation (10), the operator $\mathcal{R}_{\sigma}$ then assumes a much simpler form\footnote{If we, instead of assuming staticity, employ the Fourier transform of (5) $V(t, \mathbf{x}) = \int^{\infty}_{-\infty} \tilde{V}(\omega, \mathbf{x}) e^{- i \omega t} d \omega$ our differential operator $\mathcal{R}_{\sigma}$ can be written without loss of generality as $\mathcal{\tilde{R}}_{\sigma} = \mathcal{R}_{\sigma} + \omega^2 e^{2(\beta - \alpha)}$ effectively reducing the fully dynamical problem to that of a weighted eigenvalue problem. Thus, if one could solve the equation $\mathcal{\tilde{R}}_{\sigma} \tilde{V}(\omega, \mathbf{x}) = J^{0}$ they would have effectively solved the fully dynamical problem. Our formalism is equivalent to assuming $\omega = 0$. }
\begin{equation}
\mathcal{R}_{\sigma} := \nabla^{2} + h_{\sigma}(r) \partial_{r} = \nabla^{2} + [(1-2 \sigma) \alpha'(r) +  \beta'(r)] \partial_{r}
\end{equation}
Where the conformal metric factor is removed without any loss of generality. We will work with the operator (26) when analysing the electrostatic potentials associated with bosonic perturbations of black hole space-times. Using the ansatz expressions (21-24) we find that, after many algebraic manipulations, equation (10) reduces to:
\begin{equation}
\Big( \frac {3} {4} - \frac {(B - r)^2 (B + r)^2 [h_{\sigma}(r) (4 + r h_{\sigma}(r)) + 2 r h_{\sigma}'(r) ]} {16 B^2 r} \Big) F(\gamma) + 3 (\frac {1} {2} +  \gamma) F'(\gamma) + \gamma (1 + \gamma) F''(\gamma) = 0
\end{equation}
The above will clearly admit an ODE reduction provided the function $h_{\sigma} = (1 - 2 \sigma) \alpha'(r) + \beta'(r)$ satisfies certain conditions. For the Reissner-Nordstrom metric (4), equation (27) reduces to:
\begin{equation}
(1 - \sigma^2) F(\gamma) + 3 (\frac {1} {2} +  \gamma) F'(\gamma) + \gamma (1 + \gamma) F''(\gamma) = 0
\end{equation}
Note that the Brans-Dicke solution incorporates this case (Appendix A), but the gravitational perturbation equation $\sigma = 2$ will not be correct, since the field equations will differ.
\\
For the Reissner-Nordstrom metric we are able to find closed form solutions for a potential function $V(r,\theta)$ that is killed by $\mathcal{R}_{\sigma}$ by following the above outlined procedure. Indeed, we can now solve the ordinary differential equation (28) to obtain $F$ and consequentially $V$. 
\\
One can recognise that the above equation is relatively simple, and can in fact be solved exactly without much difficulty. We divert the reader to later sections for this exact solution and some discussion.
\\
One might attempt to see if the function $h_{\sigma}$ obeys the desired relation embedded within (27) for a range of other black holes, where we allow for a different matter Lagrangian, this will be the focus of future works.

\subsection{Conformally Flat Metrics}

We consider here applying the Hadamard method to metrics which are inherently conformally flat. Such metrics take the functional form of $g_{\mu \nu} = e^{2 \psi} \eta_{\mu \nu}$ where we allow $\psi$ to be a function of the temporal coordinate $t$ as well as the spatial ones, in light of the Birkhoff theorem no longer necessarily applying. Indeed, a wide class of metrics of this form are of interest, in particular are those of the cosmological variety. One expects the Weyl tensor to vanish for any cosmological solution, for if it does not vanish non-local properties of gravity emerge at large scales [12]. It is well known in the literature that the Friedmann-Robertson-Walker (FRW) metric is conformally flat for all the three space-time topologies it can allow [35], so it is absolutely included within our scope. Furthermore, it is of mathematical interest to examine such metrics as they admit natural Hadamard-like expansions similar to the black hole solutions in isotropic coordinates. 
\\
Note that, in the case of general relativity, $\alpha(r,t) = \beta(r,t) = \alpha(t)$ will necessarily satisfy the Friedmann-equations [35].
\\
Now, we consider $\Gamma$, which is a fundamental entity for the Hadamard expansion, to be the square of the Minkowski distance now rather than the Euclidean:
\begin{equation}
\Gamma_{M^{4}} =  -(t-t_{0})^2 + \Gamma_{\mathbb{E}^3} 
\end{equation}
Having done this, we can now expand the Hadamard series (17) in exactly the same way, noting we will have different expressions for the derivatives of $\Gamma$ and so forth. Now, if we assume that our space-time admits vanishing Weyl tensor: $C^{\alpha}_{\beta \gamma \delta} = 0$, then we can without loss of generality take $\alpha(r,t) = \beta(r,t)$, which greatly simplifies the form of equation (10). Substituting an ansatz for $V$ as the following:
\begin{equation}
V(t,r,\theta) = g(r,t) F(\Gamma_{M^{4}})
\end{equation}
where $g$ is determined uniquely by the foliation of the Hadamard series, we are able to investigate the possibility for closed form solutions to (10). Indeed, omitting the details but following along with the procedure outlined in section III, we are able to calculate that:
\begin{equation}
g(r,t) = \kappa e^{(\sigma - 1) \beta(r,t)} (t-t_{0})
\end{equation}
For any $\kappa \in \mathbb{C}$. Note the simplicity of $g$ as $\sigma \mapsto 1$, this is due to the Maxwell equations becoming particularly manageable in conformally flat space-times. Indeed, Maxwell's equations $\mathit{are}$ conformally invariant, and as such the Maxwell equations in a conformally flat space are equivalent to Maxwell's equations in Minkowski space using a Jordan frame [1]. Our results re-affirm this old known theorem, which is reassuring.
\\
As such, we find that our PDE (10) reduces to the following form:
\begin{equation}
(\sigma - 1) \big[\frac {2 \beta_{,r} + r (\beta_{,rr} + (1 - \sigma) \beta_{,r}^2 + (\sigma - 1) \beta_{,t}^2 - \beta_{,tt})} {r} \big] F(\Gamma)  + 12 F'(\Gamma) + 4 \Gamma F''(\Gamma) = 0
\end{equation}
Which will admit a unique solution provided the function $\beta$ satisfies the condition
\begin{equation}
\frac {2 \beta_{,r} + r (\beta_{,rr} + (1 - \sigma) \beta_{,r}^2 + (\sigma - 1) \beta_{,t}^2 - \beta_{,tt})} {r} =  \frac {\omega} {\sigma - 1}
\end{equation}
for some coupling constant $\omega \in \mathbb{C}$, or, of course, the other alternative where we have $\sigma = 1$. One might hope that (33) can be reduced using the Einstein or other field equations, but we leave this for future work. Nevertheless, for the case where we have $\beta(r,t) = b(t)$, our condition reduces to:
\begin{equation}
(\sigma -1) [(\sigma - 1) \dot{b}^2 - \ddot{b}] F(\Gamma) + 12 F'(\Gamma) + 4 \Gamma F''(\Gamma) = 0
\end{equation}
Indeed, working within the frame of general relativity, we are able to use the Friedmann equations explicitly. First off, we re-write our function $b(t)$ as $b(t) =\log\sqrt{a(t)}$ for simplicity. This in turn yields: 
\begin{equation}
(\sigma - 1) \big[ \frac {(\sigma +1) \dot{a}^2 - 2 a \ddot{a}} {4 a^2}\big] F(\Gamma) + 12 F'(\Gamma) + 4 \Gamma F''(\Gamma) = 0
\end{equation}
\\
Now, in the conformally flat coordinates, the Friedmann equations take the form:
\begin{equation}
\dot{a}^2 + k a^2 = \frac {8 \pi G} {3} \rho a^4
\end{equation}
\begin{equation}
\ddot{a} + ka = \frac {4 \pi G} {3} (\rho - 3p) a^3
\end{equation}
Where $k$ is the topological parameter [35]. One can now manipulate these equations to find $a$ as a function of $\rho$ and $p$ which allows for a closed form solution by demanding the additional condition that $ \frac {(\sigma +1) \dot{a}^2 - 2 a \ddot{a}} {4 a^2} = \frac {\omega} {\sigma -1}$. We find that $a$ must satisfy the algebraic equation:
\begin{equation}
\frac {k} {4} (\sigma - 1) + \frac {2 G \pi} {3} (3 p + \sigma \rho) a^2 = \frac {\omega} {\sigma -1}
\end{equation}
Which can, in principle, be readily solved for a wide variety of equations of state since this is a simple quadratic algebraic equation in terms of the function $a$. This implies that equation (32) has solution:
\begin{equation}
F(\Gamma) = \frac {1} {\Gamma \omega} [c_{1} J_{2} (\sqrt{\Gamma \omega}) + c_{2} Y_{2}(\sqrt{\Gamma \omega})]
\end{equation}
Where $J$ and $Y$ are the well known Bessel functions. 
As such, the overall solution can be immediately written down:
\begin{equation}
V_{\sigma}(t,r,\theta) = \frac {\sqrt{a(t)} e^{(\sigma - 1)} (t-t_{0})} {\Gamma \omega} [c_{1} J_{2} (\sqrt{\Gamma \omega}) + c_{2} Y_{2}(\sqrt{\Gamma \omega})]
\end{equation}
Where we have $\sigma = 0,2$. If instead we seek the electromagnetic perturbation, we find that:
\begin{equation}
V_{1}(t,r,\theta) = (t-t_{0}) [ - \frac {c_{1}} {2 \Gamma^2} + c_{2} ]
\end{equation}
As far as we know, these solutions to the perturbation equations are as yet unknown in the literature.

\section{Closed Form Solution Analysis}
Within this section we solve the relevant ODE equations which result from the method outlined in the above sections. Having solved the ODE's we then briefly discuss some of the important physical consequences of the closed form expressions to the perturbation equation (5).

\subsection{Reissner-Nordstrom Black Hole}
Recall we found the differential equation for $F$ for the Reissner-Nordstrom black hole (28):
\begin{equation}
(1 - \sigma^2) F(\gamma) + 3 (\frac {1} {2} +  \gamma) F'(\gamma) + \gamma (1 + \gamma) F''(\gamma) = 0
\end{equation}
Employing a coordinate transformation of the form $\gamma + 1 = \cosh^{2} \frac {\zeta} {2}$ the above equation reduces to:
\begin{equation}
F''(\zeta) + 2 \coth \zeta F'(\zeta) + (1 - \sigma^2) F(\zeta)  =0
\end{equation}
Which admits solution [37]:
\begin{equation}
F(\zeta) = \frac {1} {\sinh{\zeta}} (C_{1} e^{\sigma \zeta} - C_{2} e^{-\sigma \zeta}) \,\,\,\, \forall \sigma \in \mathbb{Z}
\end{equation}
Writing out the function formally now we have:
\begin{equation}
F(\gamma) = \frac {1} {2 \sqrt{\gamma} \sqrt{\gamma +1}} [C_{1} (\sqrt{\gamma +1} + \sqrt{\gamma})^{2 \sigma} - C_{2} (\sqrt{\gamma +1} - \sqrt{\gamma})^{2 \sigma}]
\end{equation}
Having found these, we can backtrack our ansatz choices and we find:
\begin{equation}
V(r,\theta) = \frac {1} {2 \eta(r) \sqrt{\gamma} \sqrt{\gamma +1}} \frac {r} {(r+B)^2} (\frac {r-B} {r+B})^{\sigma -1} [C_{1} (\sqrt{\gamma +1} + \sqrt{\gamma})^{2 \sigma} - C_{2} (\sqrt{\gamma +1} - \sqrt{\gamma})^{2 \sigma}]
\end{equation}
A rather remarkable result for the closed form solution for a bosonic perturbation. This result is in agreement with scalar and vector perturbations on the Reissner-Nordstrom black hole already present in the literature [5,26], and yields a new solution for the gravitational case.
\\
These efforts were made possible largely because the Regge-Wheeler equation is expressible in a manner very similar to the scalar and Maxwell cases after one has made some important gauge choices.

\subsection{Conformally Flat Metrics}
Recall we found that:
\begin{equation}
V_{\sigma}(t,r,\theta) = \frac {e^{(\sigma - 1) b(t)} (t-t_{0})} {\Gamma \omega} [c_{1} J_{2} (\sqrt{\Gamma \omega}) + c_{2} Y_{2}(\sqrt{\Gamma \omega})] \,\,\,\,\, \sigma = 0, \pm 2
\end{equation}
And:
\begin{equation}
V_{1}(t,r,\theta) = (t-t_{0}) [ - \frac {c_{1}} {2 \Gamma^2} + c_{2} ]
\end{equation}
Where, of course, $\Gamma = \Gamma_{M^{4}}$. We do not attempt a full scale analysis of these solutions in this paper, but we direct our efforts to a future work. However, it is very interesting to note that the solution for the electromagnetic perturbation is independent from the equation of state. This means that, when we introduce a spin 1 massless charge (i.e a photon) into a conformally flat space-time, the presence of any ordinary matter does not interfere with the dynamics of the photon. This is to be expected physically on some level, indicating that unless extreme gravitational forces are at play, the photon will not be disturbed by their presence. The wavefunction for the photon is not scattered by the potential barrier generated by the cosmological matter. This result is in agreement with other approaches [33].

\section{Discussion}
In this work we have found and expressed the forms of the differential operators describing linearised metric perturbations for arbitrary bosonic particles in general relativity. These take the form of scalar perturbations (spin 0), vector perturbations (spin 1) and tensor (gravitational) perturbations (spin 2). Having found these forms, we have used the Hadamard method as a halfway tool to find very particular series solutions to the electrostatic perturbations for black hole/stellar metrics and fully dynamical solutions for cosmological space-times wherein the Weyl tensor vanishes. Decomposing these naked series solutions we were able to reduce the original PDE equations into ordinary differential equations, which we then solved to find a general solution to the perturbation equations. 
\\
As demanded of a second order equation, we obtained two constants of integration for the partial differential equations. We have made no attempt in this paper to analyse these constants for physically meaningful boundary conditions, but we will say here that rigorous methods for their determination are present in the literature. 
In particular, using a variant of Stokes' theorem on differentiable manifolds one can express in-going and out-going radiation conditions in terms of integrals of these potential functions. These calculations have been performed in [24] for example, and we will repeat them in future works where we require particular solutions. However, for mathematically complete solutions, the details within this paper are sufficient. The forms found directly reduce to those already in the literature when we take the spin parameter $\sigma$ to equal $0$ [5] or $\sigma$ equal to $1$ [26] which gives us additional confidence in their form.

\subsection{Conclusions and Future Works}

For future work we will extend the class of metrics we wish to examine using the Hadamard method and, in particular, focus on axially symmetric metrics. Furthermore, we will analyse the potential solutions found within this paper in the next of our series. Physically meaningful data can be extracted from them in a simple way such as: values of tidal forces, plots of equipotential surfaces and the possible existence of a Meissner-effect [3,20]. Indeed, Bini et al. found that the Reissner-Nordstrom black hole admits a variant of the electric Meissner effect for spin 1 perturbations. We should like to check if this result holds for a general bosonic perturbation. Also, one can readily compute lines of force having a closed form value for the potential structure [20], this would be interesting to contrast with current Post-Newtonian expansion values.
\\
Furthermore, it goes without saying that we should like to investigate fermionic perturbations at some stage. However, given the quantum nature of fermionic perturbation equations, they must be placed on a completely different footing than bosonic ones, and as such the Hadamard method must be tweaked to permit such an extension. Nevertheless, it seems as though it should be possible for physically motivating reasons [15].
\\
In conclusion, we have successfully placed all bosonic perturbations of spin $\leq |2|$ particles on a single unifying footing, which gives hope to determine computationally efficient methods of determining tensor-wave signals.  Indeed, promising results show that this method extends to the case of axially-symmetric metrics (Kerr) as well as many from extended gravity models.

\section{References}
\small
$[1]$: K. Shankar and B. Whiting, "Self force of a static electric charge near a Schwarzschild Star", arXiv:0707.0042v2 [gr-qc] 6 Nov 2007.
\\
$[2]$: D. Bini et al, "Charged massive particle at rest in the field of a Reissner-Nordstrom black hole", arXiv:gr-qc/0609041v1 12 Sep 2006
\\
$[3]$: D. Bini et al, "On the "Electric Meissner Effect" in the field of a Reissner-Nordstrom Black Hole", Journal of the Korean Physical Society, Vol. 56, No. 5, May 2010, pp. 1594-1597
\\
$[4]$: E.T. Copson, "On Electrostatics in a Gravitational Field", Proceedings of the Royal Society of London. Series A, Containing Papers of aMathematical and Physical Character, Vol. 118, No. 779 (Mar. 1, 1928)
\\
$[5]$: A.G. Wiseman, "The self-force on a static scalar test-charge outisde a Schwarzschild black hole", arXiv:gr-qc/0001025v1 11 Jan 2000.
\\
$[6]$: P.C. Vaidya and R. Tikekar, "Exact Relativistic Model for a Superdense Star", J. Astrophys. Astr. (1982) 3, 325–334
\\
$[7]$: L.K. Patel and S.S. Koppar, "A Charged Analogue of the Viadya-Tikekar Solution, Aust. J. Phys. 1987, 40, 441-7
\\
$[8]$: R. Cai, L. Cao, "Generalized Formalism in Gauge-Invariant Gravitational Perturbations", ICTS-USTC-13-13, arXiv:1306.4927v1 [gr-qc] 20 Jun 2013
\\
$[9]$: B.C. Nolan, "Physical interpretation of gauge invariant perturbations of spherically symmetric space-times", arXiv.gr-qc/0406048v1 11 Jun 2004
\\
$[10]$: S.A. Teukolsky, "Perturbations of a rotating black hole I. Fundamental equations for gravitational, electromagnetic, and neutrino-field perturbations", The Astrophysical Journal, 185:635-647, 1973 October 15.
\\
$[11]$: T. Regge and J.A. Wheeler, "Stability of a Schwarzschild singularity", Physical R. Vol 108, No. 4, November 15, 1957.
\\
$[12]$: R.H. Brandenberger, "Lectures on the theory of cosmological perturbations", arXiv:hep-th/0306071v1 9 Jun 2003.
\\
$[13]$: K.A. Malik and D.R. Matravers, "A concise introduction to Perturbation Theory in Cosmology", Class. Quantum. Grav. 25 (2008) 193001
\\
$[14]$: M. Sasaki, H. Tagoshi, "Analytic Black Hole Perturbation Approach to Gravitational Radiation", Living Rev. Relativity, 6, (2003), 6
http://www.livingreviews.org/lrr-2003-6
\\
$[15]$: W. Tung, "Relativistic Wave Equations and Field Theory for Arbitrary Spin", Physical Review Vol. 156, No. 5, 25 April 1967.
\\
$[16]$: J. Hadamard, "Lectures on Cauchy's Problem In Linear Partial Differential Equations", New Haven: Yale University Press, Mrs. Hespa Ely Silllman Memorial Lectures, July 1921.
\\
$[17]$: M.Y. Chi, "Hadamard's Fundamental Solution and Multiple-Characteristic Problems", Rend. Sem. Mat. Univ. Pol. Torino, Vol 56 No. 3, 1998.
\\
$[18]$: H. Stephani, D. Kramer, M. A. N. MacCallum, C. Hoenselaers, E. Herlt, "Exact solu-
tions of Einsteins Field Equations" Second Edition, Cambridge Monographs on Mathematical
Physics (Cambridge University Press, 2003).
\\
$[19]$: J.M. Bardeen, W.H. Press, "Radiation fields in the Schwarzschild background", J. Math.
Phys. 14, 7 (1973).
\\
$[20]$: R. S. Hanni, R. Rufini, "Lines of force of a point charge near a Schwarzschild black
hole", Physical Review D, Vol 8 Number 10, 15 November 1973.
\\
$[21]$: P. O. Mazur, "Black Hole Uniqueness Theorems", arXiv:hep-th/0101012v1, 31 Dec 2000
\\
$[22]$: H. A. Buchdahl, "Isotropic coordinates and Schwarzschild metric", International Journal of
Theoretical Physics, Vol.24 pp. 731739, 1985
\\
$[23]$: R. Geroch et al, "A space-time calculus based on pairs of null directions", J. Math. Phys., 2005
\\
$[24]$: Maya Watanbe and Anthony W.C. Lun, "Electrostatic potential of a point charge in a Brans-Dicke Reissner-Nordstrom field", Phys. Rev. D 88, 04 5007, 2013.
\\
$[25]$: J.M. Steward and M.Walker, "Perturbations of space-times in general relativity", Proc. Roy.
Soc. Lond A341 pp.49-74, 1974
\\
$[26]$: B. Linet, "Black holes in which the electrostatic or scalar equation is solvable in closed form",
Gen. Rel. Grav. 37:2145-2163, 2005
\\
$[27]$: B. Linet, "Electrostatics and magnetostatics in the Schwarzschild metric", J. Phys. A9 pp.
1081-1087, 1976
\\
$[28]$: V. Moncrief, "Gauge-invariant perturbations of Reissner-Nordstrom black holes", Phys. Rev. D Vol 12, Issue 6, 1975.
\\
$[29]$:  J.F.Q. Fernandes and A.W.C. Lun, "Gauge invariant perturbations of black holes. I.
Schwarzschild space-time", J. Math. Phys 37 (2), 1996
\\
$[30]$:  R. Wald, "General Relativity", University of Chicago Press, 1984
\\
$[31]$: C. Misner K. Thorne and J. Wheeler, "Gravitation", W.H. Freeman and Company, San Fransicso, 1932.
\\
$[32]$: Petersen, Peter, "Riemannian geometry", Graduate Texts in Mathematics 171 (2nd ed.), Berlin, New York: Springer-Verlag, 2006.
\\
$[33]$: C.G. Tsagas, "Electromagnetic fields in curved spacetimes", arXiv:gr-qc/0407080v3 6 Jan 2005.
\\
$[34]$:  K. Giesel et al, "Manifestly Gauge-Invariant General Relativistic Perturbation Theory: II. FRW Background and First Order", arXiv:0711.0117v1 [gr-qc] 1 Nov 2007
\\
$[35]$: M. Ibison, "On the Conformal Forms of the Robertson-Walker Metric", arXiv.org/pdf/0704.2788
\\
$[36]$: K.D. Kokkotas, "Quasi-Normal Modes of Stars and Black Holes", Living Reviews in Relativity, www.livingreviews.org/Articles/Volume2/1999-2kokkotas, 16 September 1999.
\\
$[37]$: E. Kamke, Differentialgleichungen: Lsungsmethoden und
Lsungen (American Mathematical Society, Providence,
1971).

\normalsize
\section{Appendix A: Brans-Dicke Black Holes}
Taking the Brans type-I and Brans type-II metrics we write down the line element associated with the metric in isotropic coordinates as [24]:
\begin{equation}
ds^2_{BD} = -\frac {e^{2 \alpha_{0}} | \frac {r-B} {r+B} |^{\frac {2} {\lambda}}} {\eta(r)^2} + e^{2 \beta_{0}} \eta(r)^2 (1 + \frac {B} {r})^{4} | \frac {r-B} {r+B} |^{2 \frac {(\lambda - C - 1)} {\lambda}} (dr^2 + r^2 d \Omega^2)
\end{equation}
Where we have:
\begin{equation}
\eta(r) = p_{+}^2 - p_{-}^2 | \frac {r-B} {r+B} |^{2 k}
\end{equation}
The $p_{\pm}^2$ constitute the horizons of the black hole similar to the Reissner-Nordstrom case, we have set the parameter $k = \frac {C+2} {2 \lambda}$ and the scalar field for the theory assumes the form $\phi = \phi_{0} | \frac {r-B} {r+B} |^{\frac {C} {\lambda}}$.
\\
Writing $\alpha$ and $\beta$ as per equation (1) we get, following the methods described by employing the ansatz equations (21-24), the ODE:
\begin{equation}
\frac{ -(C-2 \lambda + 2 \sigma)(C + 2 \lambda + 2 \sigma) F(\gamma)} {4 \lambda^2} + 3 (\frac {1} {2} +  \gamma) F'(\gamma) + \gamma (1 + \gamma) F''(\gamma) = 0
\end{equation}
Which can readily be solved to in turn yield an expression for the scalar and electrostatic potentials associated with a spin 0 and spin 1 perturbation of the Brans-Dicke space-time. We omit the solution here as it is lengthy and contains hypergeometric function expressions.

\section{Appendix B: Photon Perturbation Of Stellar Interior}
We now take to examining the Stellar interior solution as mentioned earlier in the introduction. We do this not only for completeness but to demonstrate the power of the Hadamard-Copson method. Assume the background metric (6) for this section.
\\
Employing the ansatz $(21-23)$ and the choice of $f$ in $(25)$ where we pick the constants as:
\begin{equation}
\mu = \frac {-\alpha} {\iota} \,\, ; \kappa = \frac {\beta} {\sqrt{\iota}} \,\, ; \iota = \sqrt{2 \alpha} \beta
\end{equation}
Evaluating equation (10), we find:
\begin{equation}
\mathcal{R}_{1} V = 3(\gamma -1) F'(\gamma) + (\gamma - 2) \gamma F''(\gamma) = 0
\end{equation}
Which can be reduced to a first order ODE by letting $F'(\gamma) = G(\gamma)$ and hence we have the solution:
\begin{equation}
F(\gamma) = C_{2} + \frac {C_{1} (\gamma -1)} {\gamma \sqrt{2 - \gamma}}
\end{equation}
Which yields an interesting contrast with the other solutions. Substituting everything back in, we find that the electrostatic potential $V$ can be written:
\begin{equation}
V(r,\theta) =  C_{2} + \frac {C_{1} (\frac {\Gamma_{\mathbf{E}^3}} {k(r) k(b)} -1)} {\frac {\Gamma_{\mathbf{E}^3}} {k(r) k(b)} \sqrt{2 - \frac {\Gamma_{\mathbf{E}^3}} {k(r) k(b)}}}
\end{equation}
Where we have:
\begin{equation}
k(r) = \frac {\mu r^2 - \kappa^2} {\iota}
\end{equation}
The expression immediately demands that $\beta$ be negative, which is not an issue since only $\beta^2$ appears within the metric (6). This result is a direct generalisation of [1], and in fact implies that we have a summation formula for the multipole series obtained therein.

\end{document}